\documentclass[useAMS,usenatbib]{mn2e}

\title[Analysing Astronomy Algorithms for GPUs and Beyond]
	  {Analysing Astronomy Algorithms for GPUs and Beyond}
\author[B. R. Barsdell et al.]
  {B. R. Barsdell,$^1$\thanks{Corresponding author: bbarsdel@astro.swin.edu.au}
  D. G. Barnes,$^1$
  C. J. Fluke$^1$
  \newauthor 
  \\
  $^1$Centre for Astrophysics and Supercomputing, Swinburne University of
  Technology,\\
  PO Box 218, Hawthorn, Australia, 3122}

\pagerange{\pageref{firstpage}--\pageref{lastpage}} \pubyear{2010}

\def\LaTeX{L\kern-.36em\raise.3ex\hbox{a}\kern-.15em
    T\kern-.1667em\lower.7ex\hbox{E}\kern-.125emX}

\usepackage{graphicx}
\usepackage{url}

\begin{document}

\label{firstpage}

\maketitle

\begin{abstract}
Astronomy depends on ever increasing computing power.
Processor clock-rates have plateaued, and increased performance is now
appearing in the form of additional processor cores on a single chip. This
poses significant challenges to the astronomy software community.
Graphics Processing Units (GPUs), now capable of general-purpose computation,
exemplify both the difficult learning-curve and the significant speedups
exhibited by massively-parallel hardware architectures. We present a
generalised approach to tackling this paradigm shift, based on the analysis of
algorithms.
We describe a small collection of foundation algorithms relevant to astronomy
and explain how they may be used to ease the transition to massively-parallel
computing architectures. We demonstrate the effectiveness of our approach by
applying it to four well-known astronomy problems: H\"ogbom \textsc{clean},
inverse ray-shooting for gravitational lensing, pulsar dedispersion and
volume rendering.
Algorithms with well-defined memory access patterns and high arithmetic
intensity stand to receive the greatest performance boost from
massively-parallel architectures, while those that involve a significant
amount of decision-making may struggle to take advantage of the available
processing power.
\end{abstract}


\section{Introduction}
Computing resources are a fundamental tool in astronomy: they
are used to acquire and reduce observational data, simulate
astrophysical processes, and analyse and visualise the results.
Advances in the field of astronomy have depended heavily on the
increase in computing power that has followed Moore's
Law \citep{moore1965} since the mid 1960s; indeed, many
contemporary astronomy survey projects and astrophysics simulations
would simply not be possible without the evolution Moore predicted.

Until recently, increased computing power was delivered in direct
proportion to the increase in central processing unit (CPU) clock
rates. Astronomy software executed more and more rapidly
with each new hardware release, without any further programming work.
But around 2005, the advance in clock rates ceased, and manufacturers
turned to increasing the {\em instantaneous}\/ processing capacity of
their CPUs by including additional processing cores in a
single silicon chip package.  Today's mainstream multi-core
CPUs typically have between 2 and 8 processing cores; these are routinely
deployed in large-scale compute clusters.

\begin{figure}
\includegraphics[width=8.5cm]{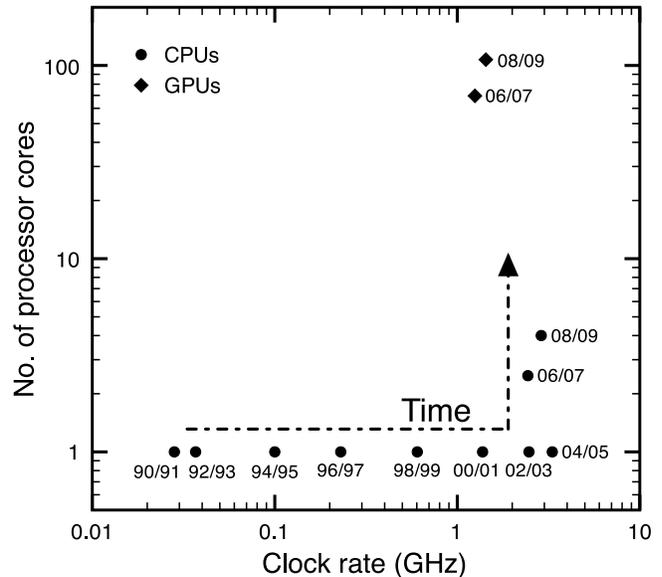} 
\vspace{-20pt}
\caption{Clock-rate versus core-count phase space of Moore's Law binned every
  two years for CPUs (circles) and GPUs (diamonds). There is a general trend
  for performance to increase from bottom left to top right.}
\label{fig:MulticoreMooresLaw}
\end{figure}

Fig.~\ref{fig:MulticoreMooresLaw} places the mainstream CPUs from
the last $\sim 20$ years in the clock rate versus core-count phase
space. In this space, the evolution of CPUs turns a `corner' around 2005
when clock rates plateaued and multi-core processors emerged. Lying
above today's fastest multi-core CPUs in
Fig.~\ref{fig:MulticoreMooresLaw} though, are the contemporary
graphics processing units (GPUs), boasting hundreds of cores (`{\em
  many-core}') and $\sim 1$~Ghz clock rates.  GPUs are already useful
in their own right---providing $\sim 30$ times the
raw computation speed of CPUs---but perhaps more interestingly, they
represent the likely 
evolution of CPUs. GPUs demonstrate how computing power can continue to follow
Moore's Law in an era of zero (or even negative) growth in clock rates.

The plateau in processor clock rates is problematic for astronomy
software composed of {\em sequential}\/ codes, wherein instructions are
executed one after the other.
Such codes derive no
direct performance benefit from the presence of multiple processing
cores, and their performance will languish for as long as processor
clock rates remain steady at $\sim3$--4~GHz. Astronomy software must be
(re-)written to take advantage of many-core processors.

Astronomers are already cogniscent of this issue. Shared- and
distributed-memory multi-core CPU machines have been exploited using the
well-known OpenMP and MPI programming models\footnote{While the specifics of
  parallel CPU systems lie outside the scope of 
  this paper, we note that many of the algorithm analysis techniques we
  describe lend themselves equally well to these architectures.}.
In addition, a number of researchers have
adapted, written and/or re-written classic astronomy codes for the GPU
architecture in the last $\sim3$ years, and gained performance
improvements ranging from factors of a few to factors of several
hundred. Some highlights include N-body (e.g.,
\citealt{HamadaEtal2009}), radio-telescope signal correlation (e.g.,
\citealt{WaythEtal2009}), adaptive mesh refinement (e.g.,
\citealt{SchiveEtal2010}), galaxy spectral energy distribution
\citep{JonssonPrimack2009} and gravitational microlensing
\citep{ThompsonEtal2010} codes.

Inevitably, a section of the
astronomy community will continue with an {\em ad hoc}\/ approach to
the adaptation of software from single-core to many-core
architectures. 
In this paper, we demonstrate that there is a significant difference
between current computing techniques and those required to efficiently
utilise new hardware architectures such as many-core processors, as
exemplified by GPUs.  These techniques will be unfamiliar to most
astronomers and will pose a challenge in terms of keeping our
discipline at the forefront of computational science.  We present a
practical, effective and simple methodology for creating astronomy
software whose performance scales well to present and future many-core
architectures.  Our methodology is grounded in the classical computer
science field of algorithm analysis.

In Section~\ref{sec:ourmethodology} we introduce the key concepts in
algorithm analysis, with particular focus on the context of many-core
architectures. We present four foundation algorithms, and characterise them
as we outline our algorithm analysis methodology. In Section
\ref{sec:AstronomyAlgorithms} we demonstrate the proposed methodology by
applying it to four well-known astronomy problems, which we break down into
their constituent foundation algorithms. We validate our analysis of these
problems against {\em ad hoc}\/ many-core implementations as available in the
literature and discuss the implications of our approach for the future of
computing in astronomy in Section \ref{sec:discussion}.

\section{A Strategic Approach: Algorithm Analysis}
\label{sec:ourmethodology}

Algorithm analysis, pioneered by Donald Knuth (see, e.g.,
\citealt{Knuth1998}), is a fundamental component of computer science -- a 
discipline that is more about how to solve problems than the actual
implementation in code.
In this work, we are not interested in the specifics (i.e., syntax) of
implementing a given astronomy algorithm with a particular programming
language or library (e.g., CUDA, OpenCL, Thrust) on a chosen computing
architecture (e.g., GPU, Cell, FPGA). As \citet{Harris2007} notes,
algorithm-level optimisations are much more important with respect to overall
performance on many-core hardware (specifically GPUs) than implementation
optimisations, and should be made first. We will return to the issue of
implementation in future work.

Here we present an approach to tackling the transition to many-core hardware
based on the analysis of algorithms. The purpose of this analysis is to
determine the potential of a given algorithm for a many-core architecture
\textit{before} any code is written. This provides essential information about
the optimal approach as well as the return on investment one might expect for
the effort of (re-)implementing a particular algorithm. Our methodology was in
part inspired by the work of \citet{Harris2005}.

Work in a similar vein has also been
undertaken by \citet{AsanovicEtal2006, AsanovicEtal2009} who classified
parallel algorithms into 12 groups, referring to them as `dwarfs'.
While insightful and opportune, these 
dwarfs consider a wide range of parallel architectures, cover all areas of
computation (including several that are not of great relevance to
astronomy) and are limited as a resource by the coarse nature of the
classification. In contrast, the approach presented here is tailored to the 
parallelism offered by many-core processor architectures, contains algorithms
that appear frequently within astronomy computations, and provides a
fine-grained level of detail. Furthermore, our approach considers the
fundamental concerns raised by many-core architectures at a level of
abstraction that avoids dealing with hardware or software-specific details and
terminology. This is in contrast to the work by \citet{CheEtal2008}, who
presented a useful but highly-targeted summary of general-purpose programming
on the NVIDIA GPU architecture.

For these reasons this work will serve as a valuable and practical resource
for those wishing to analyse the expected performance of particular astronomy
algorithms on current and future many-core architectures.

For a given astronomy problem, our methodology is as follows:
\begin{enumerate}
  \item Outline each step in the problem.
  \item Identify steps that resemble known algorithms (see below).
  \begin{enumerate}
    \item Outlined steps may need to be further decomposed into sub-steps before
    a known counterpart is recognised. Such composite steps
    may later be added to the collection of known algorithms.
  \end{enumerate}
  \item For each identified algorithm, refer to its pre-existing analysis.
  \begin{enumerate}
    \item Where a particular step does not appear to match any known algorithm,
	  refer to a relevant analysis methodology to analyse the step
      as a custom algorithm (see Sections \ref{sec:CriticalIssues},
      \ref{sec:complexity_analysis} and \ref{sec:analysis}). The
      newly-analysed algorithm can then be added to the collection
	  for future reference.
  \end{enumerate}
  \item Once analysis results have been obtained for each step, apply
  a global analysis to the algorithm to obtain a complete picture of its
  behaviour (see Section \ref{sec:macro_analysis}).
\end{enumerate}

Here we present a
small collection of foundation algorithms\footnote{Note that for
  these algorithms we have used naming conventions that are familiar to us but
  are by no means unique in the literature.} that appear in
computational astronomy problems. This is motivated by the fact that
complex algorithms may be \textit{composed} from simpler ones. We propose that
\textit{algorithm composition} provides an excellent approach to turning the
multi-core corner. Here we focus on its application to algorithm analysis; in
future work we will show how it may also be applied to implementation
methodologies.
The algorithms are described below using a \textit{vector} data
structure. This
is a data structure like a Fortran or C array representing a contiguous block
of memory and providing constant-time random access to individual
elements\footnote{Here we use \textit{constant-time} in the algorithmic sense,
  i.e., constant with respect to the size of the input data. In this context
  we are not concerned with hardware-specific performance factors.}.
We use the notation $v[i]$ to represent the $i^{\rm th}$ element of a
vector $v$.

 \textbf{Transform:} Returns a vector containing the result of the
application of a specified function to every individual element of an input
vector.
\begin{equation}
{\rm out}[i] = f({\rm in}[i])
\end{equation}
Functions of more than one variable may also be applied to multiple input
vectors. Scaling the brightness of an image (defined as a vector of pixels) is
an example of a transform operation.

 \textbf{Reduce:} Returns the sum of every element in a vector.
\begin{equation}
{\rm out} = \sum_i {\rm in}[i]
\end{equation}
Reductions may be generalised to use any associative binary operator, e.g.,
product, min, max etc. Calculating image noise is a common application of the
reduce algorithm.

 \textbf{Gather:} Retrieves values from an input vector according to a
  specified index mapping and writes them to an output vector.
\begin{equation}
{\rm out}[i] = {\rm in}[{\rm map}[i]]
\end{equation}
Reading a shifted or transformed subregion of an image is a common example
of a gather operation.

 \textbf{Interact:} For each element $i$ of an input vector, in$_1$, sums the
  interaction between $i$ and each element $j$ in a second input vector,
  in$_2$.
\begin{equation}
{\rm out}[i] = \sum_j f({\rm in}_1[i], {\rm in}_2[j])
\end{equation}
where $f$ is a given interaction function. The best-known application of this
  algorithm in astronomy is the computation of forces in a direct N-body
  simulation, where both input vectors represent the system's particles and
  the interaction function calculates the gravitational force between two
  particles.

These four algorithms were chosen from experience with a number
of computational astronomy problems.
The transform, reduce and gather operations may be referred to as `atoms' in
the sense that they are indivisible operations. While the interact algorithm
is technically a composition of transforms and reductions, it will be analysed
as if it too was an atom, enabling rapid analysis of problems that use the
interact algorithm without the need for further decomposition.

We now describe a number of algorithm analysis techniques that we
have found to be relevant to massively-parallel architectures. These
techniques should be applied to the individual algorithms that comprise a
complete problem in order to gain a detailed understanding of their
behaviour.

\subsection{Principle characteristics}
\label{sec:CriticalIssues}
Many-core architectures exhibit a number of characteristics that can impact
strongly on the performance of an algorithm. Here we summarise four of the
most important issues that must be considered.

\textbf{Massive parallelism:} To fully utilise massively-parallel
architectures, algorithms must exhibit a 
high level of parallel \textit{granularity}, i.e., the number of required
operations that may be performed simultaneously must be large and
scalable. \textit{Data-parallel} algorithms, which divide their \textit{data} 
between parallel processors rather than (or in addition to) their
\textit{tasks}, exhibit parallelism that scales with the size of their input
data, making them ideal candidates for massively-parallel
architectures. However, performance may suffer when these algorithms are
executed on sets of input data that are small relative to the number of
processors in a particular many-core architecture\footnote{Note also that
  oversubscription of threads to processors is often a requirement for good
  performance in many-core architectures. For example, an NVIDIA GT200-class
  GPU may be under-utilised with an allocation of fewer than $\sim 10^4$
parallel threads, corresponding to an oversubscription rate of around
$50\times$.}.

\begin{figure}
\includegraphics[width=8cm]{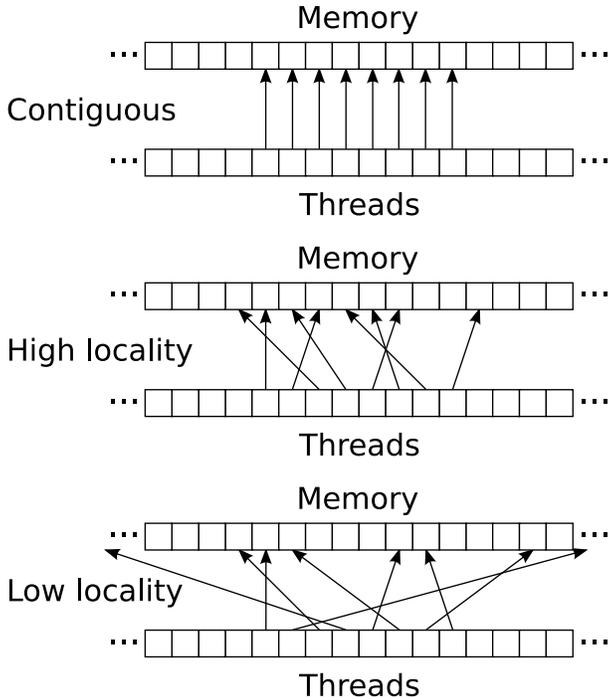} 
\caption{Representative memory access patterns indicating varying levels of
  locality of reference.
  Contiguous memory access is the optimal case for many-core
  architectures. Patterns with high locality will generally achieve good
  performance; those with low locality may incur severe performance
  penalties.}
\label{fig:memory_access_patterns}
\end{figure}

\textbf{Memory access patterns:} Many-core architectures contain very
high bandwidth main memory\footnote{Memory bandwidths on current GPUs are
 $\mathcal{O}(100$GB/s$)$.} 
in order to `feed' the large number of parallel processing
units. However, high latency (i.e., memory transfer startup) costs mean that
performance depends strongly on the \textit{pattern} in which memory is
accessed.
In general, maintaining `locality of reference' (i.e., neighbouring threads
accessing similar locations in memory) is vital to achieving good
performance\footnote{Locality of reference also affects performance on traditional
 CPU architectures, but to a lesser extent than on GPUs.}. Fig.
\ref{fig:memory_access_patterns} illustrates different levels of locality of
reference.

Collisions between threads trying to read the same location in
memory can also be costly, and write-collisions must be treated using
expensive atomic operations in order to avoid conflicts between threads.

\begin{figure}
\includegraphics[width=8cm]{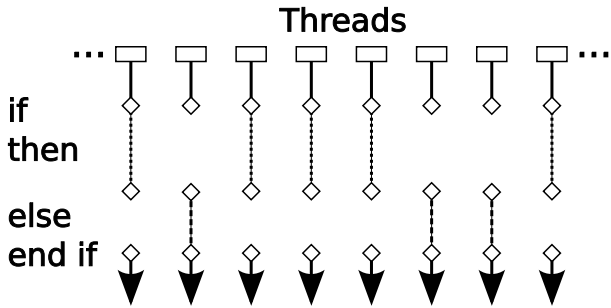} 
\caption{A schematic view of divergent execution within a SIMD
  architecture. Lines indicate the 
  flow of instructions; white diamonds indicate branch points, where the 
  code paths of neighbouring threads diverge. The statements on the left
  indicate typical corresponding source code. White space between branch
  points indicates a thread waiting for its neighbours to complete a divergent
  code section.}
\label{fig:branching}
\end{figure}

\textbf{Branching:} Current many-core architectures 
rely on single instruction multiple data (SIMD) hardware.
This means that neighbouring threads 
that wish to execute different instructions must
wait for each other to complete the divergent code section before execution
can continue in parallel (see Fig. \ref{fig:branching}).
For this reason, algorithms that involve significant branching between
different threads may suffer severe performance degradation. Similar to the
effects of memory access locality, performance will in general depend on the
locality of branching, i.e., the number of different code-paths taken by a
group of neighbouring threads.

\textbf{Arithmetic intensity:} Executing arithmetic instructions is generally
much faster than accessing memory on current many-core hardware. Algorithms
performing few arithmetic operations per memory access may become
memory-bandwidth-bound; i.e., their speed becomes limited by the rate at
which memory can be accessed, rather than the rate at which arithmetic
instructions can be processed. Memory bandwidths in many-core architectures
are typically significantly higher than in CPUs, meaning that even
bandwidth-bound algorithms may exhibit strong performance; however, they will 
not be able to take full advantage of the available computing
power.
In some cases, it may be
beneficial to re-work an algorithm entirely in order to increase its
arithmetic intensity, even at the cost of performing more numerical work in
total.

For the arithmetic intensities presented in this
paper, we assume an idealised cache model in which only the first memory read
of a particular piece of data is included in the count; subsequent or parallel
reads of the same data are assumed to be made from a cache, and are not
counted. The ability to achieve this behaviour in practice will depend
strongly on the memory access pattern (specifically the locality of memory
accesses).

\begin{table*}
\caption{Analysis of four foundation algorithms}
\begin{tabular}{lllllllll}
\hline
\label{tbl:analysis_of_simple_algorithms}
                       & Transform        & Reduction                     & Gather           & Interact                                 \\
\hline
Work                   & $\mathcal{O}(N)$ & $\mathcal{O}(N)$              & $\mathcal{O}(N)$ & $\mathcal{O}(NM)$                        \\
Depth                  & $\mathcal{O}(1)$ & $\mathcal{O}(\log N)$         & $\mathcal{O}(1)$ & $\mathcal{O}(M)$ or $\mathcal{O}(\log M)$\\
Memory access locality & Contiguous       & Contiguous                    & Variable         & Contiguous                               \\
Arithmetic intensity   & $1:1:\alpha$     & $1:\frac{1}{N}:\alpha$        & $1:1:0$          & $1+\frac{M}{N}:1:2M\alpha$ \\
\hline
\end{tabular}
\end{table*}

\subsection{Complexity analysis}
\label{sec:complexity_analysis}
The complexity of an algorithm is a formal measure of its execution time
given a certain size of input. It is often used as a means of comparing the
speeds of two different algorithms that compute the same (or a similar)
result. Such comparisons are critical to understanding the relative
contributions of different parts of a composite algorithm and identifying
bottle-necks.

Computational complexity is typically expressed as the 
total run-time, $T$, of an algorithm as a function of the input size, $N$,
using `Big O' notation.
Thus $T(N) = \mathcal{O}(N)$ means a run-time that is 
proportional to the input size $N$. An algorithm with complexity of $T(N) =
\mathcal{O}(N^2)$ will take four times as long to run after a doubling of its
input size.

While the complexity measure is traditionally used for algorithms running on
serial processors, it can be generalised to analyse parallel algorithms. One
method is to introduce a second parameter: $P$, the number of processors. The
run-time is then expressed as a function of both $N$ and $P$. For example, an
algorithm with a parallel complexity of $T(N, P) = \mathcal{O}(\frac{N}{P})$ will run
$P$ times faster on $P$ processors than on a single processor for a given
input size; i.e., it exhibits perfect parallel scaling. More complex
algorithms may incur overheads when run in parallel, e.g., those requiring
communication between processors. In these cases, the parallel complexity 
will depend on the specifics of the target hardware architecture.

An alternative way to express parallel complexity is using the \textit{work},
$W$, and \textit{depth}, $D$, metrics first introduced formally by
\citet{Blelloch1996}. Here, work measures the total number of computational 
operations performed by an algorithm (or, equivalently, the run-time on a
single processor), while depth measures the longest
sequence of sequentially-dependent operations (or, equivalently, the run-time
on an infinite number of processors). The depth metric is a measure of the
amount of inherent parallelism in the algorithm.
A perfectly parallel algorithm has work complexity of $W(N) =  \mathcal{O}(N)$ and depth
complexity of $D(N) = \mathcal{O}(1)$, meaning all but a constant number of operations
may be performed in parallel.
An algorithm with $W = \mathcal{O}(N)$ and $D = \mathcal{O}(\log{N})$ is highly parallel, but
contains some serial dependencies between operations that scale as a function
of the input size.
Parallel algorithms with work complexities equal to those of their serial
counterparts are said to be `work efficient'; those that further exhibit low
depth complexities are considered to be efficient parallel algorithms. The
benefit of the work/depth metrics over the parallel run-time is that they have
no dependence on the particular parallel architecture on which the algorithm
is executed, i.e., they measure properties inherent to the algorithm.

A final consideration regarding parallel algorithms is Amdahl's law
\citep{Amdahl1967}, which states that the maximum possible speedup over a
serial algorithm is limited by the fraction of the parallel algorithm that
cannot be (or simply \textit{is not}) parallelised. Assuming an infinite
number of available processors, the run-time of the parallel part of the
algorithm will reduce to a constant, while the serial part will continue to
scale with the size of the input. In terms of the work/depth metrics, the
depth of the algorithm represents the fraction that cannot be parallelised,
and the maximum theoretical speedup is given by $S_\mathrm{max} \approx
\frac{W}{D}$. Note the implication that the maximum speedup is actually a
function of the input size. Increasing the problem size in addition to the
number of processors allows the speedup to scale more effectively.

\subsection{Analysis results}
\label{sec:analysis}
We have applied the techniques discussed in Sections
\ref{sec:CriticalIssues} and \ref{sec:complexity_analysis} to the 
four foundation algorithms introduced at the beginning of Section
\ref{sec:ourmethodology}. 
We use the following metrics:
\begin{itemize}
\item \textbf{Work and depth:} The complexity metrics as described in Section
 \ref{sec:complexity_analysis}.
\item \textbf{Memory access locality:} The nature of the memory access
  patterns as discussed in Section \ref{sec:CriticalIssues}.
\item \textbf{Arithmetic intensity:} Defined by the triple ratio $r:w:f$
  representing the number of read, write and function evalation operations
  respectively that the algorithm performs (normalised to the input size). The
  symbol $\alpha$ is used, where applicable, to represent the internal
  arithmetic intensity of the function given to the algorithm.
\end{itemize}
The results are presented in Table
  \ref{tbl:analysis_of_simple_algorithms}. Note that this analysis is based on 
  the most-efficient known parallel version of each algorithm.

\subsection{Global analysis}
\label{sec:macro_analysis}
Once local analysis results have been obtained for each step of a problem, it
is necessary to put them together and perform a global analysis. Our
methodology is as follows:
\begin{enumerate}
\item Determine 
  the components of the algorithm where most of the computational work lies by
  comparing work complexities. Components with similar work complexities
  should receive similar attention with respect to parallelisation in order to
  avoid leaving behind bottle-necks as a result of Amdahl's Law.
\item Consider the amount of inherent parallelism in each algorithm

by observing its theoretical speedup $S_{\rm max} \approx \frac{W}{D}$.
\item Use the theoretical arithmetic intensity of each algorithm to
  determine the likelihood of it being limited by memory bandwidth rather than
  instruction throughput. The theoretical \textit{global} arithmetic intensity
  may be obtained by comparing the total amount of input and output data to
  the total amount of arithmetic work to be done in the problem.
\item Assess the memory access patterns of each algorithm to identify the
  potential to achieve peak arithmetic intensity\footnote{Studying the memory access
  patterns will also help to identify the optimal caching strategy if this
  level of optimisation is desired.}.
\item If particular components exhibit poor properties, consider alternative
  algorithms.
\item Once a set of component algorithms with good theoretical performance
  has been obtained, the algorithm decomposition
should provide a good starting point for an implementation.
\end{enumerate}

\section{Application to Astronomy Algorithms}
\label{sec:AstronomyAlgorithms}
We now apply our methodology from Section \ref{sec:ourmethodology} to four
typical astronomy computations. In each case, we demonstrate how to identify
the steps in an outline of the problem as foundation algorithms from our
collection described at the beginning of Section \ref{sec:ourmethodology}. We
then use this knowledge to study the exact nature of the available parallelism
and determine the problem's overall suitability for many-core
architectures.
We note that we have deliberately chosen simple versions of the problems in
order to maximise clarity and brevity in illustrating the principles of our
algorithm analysis methodology.

\subsection{Inverse ray-shooting gravitational lensing}
\label{sec:rayshooting}
\textbf{Introduction:} Inverse ray-shooting is a numerical technique used in
gravitational microlensing. Light rays are projected backwards (i.e., from
the observer) through an ensemble of lenses and on to a source-plane pixel 
grid. The number of rays that fall into each pixel 
gives an indication of the magnification at that spatial position 
relative to the case where there was no microlensing. In cosmological
scenarios, the resultant maps are used to study brightness variations 
in light curves of lensed quasars, providing constraints on the physical
size of the accretion disk and broad line emission regions.

The two main approaches to ray-shooting are based on either the direct
calculation of the gravitational deflection by each lens
\citep{KayserEtal1986, SchneiderWeiss1986, SchneiderWeiss1987} or the use of a 
tree hierarchy of psuedo-lenses \citep{Wambsganss1990, Wambsganss1999}. Here,
we consider the direct method.

\noindent \textbf{Outline:} The ray-shooting algorithm is easily divided
into a number of distinct steps: 

\begin{enumerate}
\item Obtain a collection of lenses according to a desired
distribution, where each lens has position and mass.
\item Generate a collection of rays according to a uniform
  distribution within a specified 2D region, where each ray is defined by its
  position.
\item For each ray, calculate and sum its deflection due to each lens.
\item Add each ray's calculated deflection to its initial position to obtain
  its deflected position.
\item Calculate the index of the pixel that each ray falls into.
\item Count the number of rays that fall into each pixel.
\item Output the list of pixels as the magnification map.
\end{enumerate}

\noindent \textbf{Analysis:} To begin the analysis, we interpret the above
outline as follows:

\begin{itemize}
\item Steps (i) and (ii) may be considered \textit{transform} operations that
  initialise the vectors of lenses and rays.
\item Step (iii) is an example of the \textit{interact} algorithm, where
the inputs are the vectors of rays and lenses and the interaction function
calculates the deflection of a ray due to the gravitational potential around a
lens mass.
\item Steps (iv) and (v) apply further transforms to the collection of
  rays.
\item Step (vi) involves the generation of a histogram.
As we have not already identified this algorithm in Section
\ref{sec:ourmethodology},
it will be necessary to analyse this step as a unique algorithm.
\end{itemize}

According to this analysis, three basic algorithms
comprise the complete technique: transform, interact and
histogram generation. Referring to Table
\ref{tbl:analysis_of_simple_algorithms}, we see that, in the context of a
lensing simulation using $N_{\rm rays}$ rays and $N_{\rm lenses}$ lenses, the
amount of work performed by the transform and interact algorithms will be $W =
\mathcal{O}(N_{\rm rays}) + \mathcal{O}(N_{\rm lenses})$ and $W =
\mathcal{O}(N_{\rm rays} N_{\rm lenses})$ respectively.

We now analyse the histogram step. Considering
first a serial algorithm for generating a 
histogram, where each point is considered in turn and the
count in its corresponding bin is incremented, we find the
work complexity to be $W = \mathcal{O}(N_{\rm rays})$. Without further analysis, we
compare this to those of the other component algorithms. The serial histogram
and the transform operations each perform similar work. The interact algorithm
on the other hand must, as we have seen, perform work proportional to $N_{\rm
  rays} \times N_{\rm lenses}$. 
For large $N_{\rm lenses}$ (e.g., as occurs in cosmological microlensing
simulations, where $N_{\rm lenses} > 10^4$) this step will dominate the total
work. Assuming the number of lenses is scaled with the amount of parallel
hardware,
the interact step will also dominate the total run-time.

Given the dominance of the interact step, we now choose
to ignore the effects of the other steps in the problem. It
should be noted, however, that in contrast to cosmological microlensing,
planetary microlensing models contain only a few lenses. In this case, the
work performed by the interact step will be similar to that of the other
steps, and thus the use of a serial histogram algorithm alongside parallel
versions of all other steps would result in a severe performance
bottle-neck. Several parallel histogram algorithms exist, but a discussion of
them is beyond the scope of this work.

Returning to the analysis of the interact algorithm, we
again refer to Table \ref{tbl:analysis_of_simple_algorithms}. Its worst-case
depth complexity indicates a maximum speedup of $S \approx W =
\mathcal{O}(N_{\rm rays})$, i.e., parallel speedup scaling perfectly up to the
number of rays. The 
arithmetic intensity of the algorithm scales as $N_{\rm lenses}$ and will thus
be very high. Contiguous memory accesses indicate strong potential to achieve
this high arithmetic intensity.
We conclude that direct inverse ray-shooting for cosmological microlensing is
an ideal candidate for an efficient implementation on a many-core
architecture.

\subsection{H\"ogbom CLEAN}
\textbf{Introduction:} Raw (`dirty') images produced by radio interferometers
exhibit unwanted artefacts as the result of the incomplete sampling of the
visibility plane. These artefacts can inhibit image analysis and should
ideally be removed by deconvolution.
Several different techniques have been developed to `clean' these
images. For a review, see \citet{Briggs1995}. Here we analyse the image-based
algorithm first described by \citet{Hogbom1974}. We note that the algorithm by
\citet{Clark1980} is now the more popular choice in the astronomy community,
but point out that it is essentially an approximation to H\"ogbom's algorithm
that provides increased performance at the cost of reduced accuracy.

The algorithm involves iteratively finding the brightest point in the `dirty
image' and subtracting from the dirty image an image of the beam centred on
and scaled by this brightest point. The procedure continues until the
brightest point in the image falls below a prescribed threshold. 
While the iterative procedure must be performed sequentially, the computations
within each iteration step are performed independently for every pixel of the
images, suggesting a substantial level of parallelism.
The output of the algorithm is a series of `clean components', which may be
used to reconstruct a cleaned image.

\noindent \textbf{Outline:} The algorithm may be divided into the following
steps: 
\begin{enumerate}
\item Obtain the beam image.
\item Obtain the image to be cleaned.
\item Find the brightest point, $b$, the standard deviation, $\sigma$, and the
  mean, $\mu$, of the image.
\item If the brightness of $b$ is less than a prescribed threshold (e.g.,
  $|b-\mu|<3\sigma$), go to step (ix).
\item Scale the beam image by a fraction (referred to as the `loop gain') of
  the brightness of $b$.
\item Shift the beam image to centre it over $b$.
\item Subtract the scaled, shifted beam image from the input image to produce
  a partially-cleaned image.
\item Repeat from step (iii).
\item Output the `clean components'.
\end{enumerate}

\noindent \textbf{Analysis:} We decompose the outline of the H\"ogbom
\textsc{clean} algorithm as follows:
\begin{itemize}
\item Steps (i) and (ii) are simple data-loading operations, and may be
  thought of as transforms.
\item Step (iii) involves a number of reduce operations over the pixels in the
  dirty image.
\item Step (v) is a transform operation, where each pixel in the beam is
  multiplied by a scale factor.
\item Step (vi) may be achieved in two ways, either by directly reading an
  offset subset of the beam pixels, or by switching to the Fourier domain and
  exploiting the shift theorem. Here we will only consider the former option,
  which we identify as a gather operation.
\item Step (vii) is a transform operation over pixels in the dirty image.
\end{itemize}

We thus identify three basic algorithms in H\"ogbom \textsc{clean}:
\textit{transform}, \textit{reduce} and \textit{gather}. Table
\ref{tbl:analysis_of_simple_algorithms} shows that the work performed by each
of these algorithms will be comparable (assuming the input and beam images are
of similar pixel resolutions). This suggests that any acceleration should be
applied equally to \textit{all} of the steps in order to avoid the creation of
bottle-necks.

The depth complexities of each algorithm indicate a limiting speed-up of
$S_{\rm max} \approx \mathcal{O}(\frac{N_{\rm pxls}}{\log{N_{\rm pxls}}})$ during the
reduce operations. While not quite ideal, this is still a good
result. Further, the algorithms do not exhibit high arithmetic intensity (the
calculations involving only a few subtractions and multiplies) and are thus
likely to be bandwidth-bound. This will dominate any effect the limiting
speed-up may have.

The efficiency with which the algorithm will use the available memory
bandwidth will depend on the memory access patterns. The transform and reduce
algorithms both make contiguous memory accesses, and will thus achieve peak
bandwidth. The gather operation in step (vi), where the beam image is shifted
to centre it on a point in the input image, will access memory in an offset
but contiguous 2-dimensional block. This 2D locality suggests the
potential to achieve near-peak memory throughput.

We conclude that the H\"ogbom \textsc{clean} algorithm represents a good candidate for
implementation on many-core hardware, but will likely be bound by the
available memory bandwidth rather than arithmetic computing performance.

\subsection{Volume rendering}
\textbf{Introduction:} There are a number of sources of volume data in
astronomy, including spectral cubes from radio telescopes and integral field
units, as well as simulations using adaptive mesh refinement and smoothed
particle hydrodynamics techniques. Visualising these data in
physically-meaningful ways is important as an analysis tool, but even small
volumes (e.g., $256^3$) require large amounts of computing power to render,
particularly when real-time interactivity is desired.

Several methods exist for rendering volume data; here we analyse a direct (or
\textit{brute-force}) ray-casting algorithm \citep{Levoy1990}. While
similarities exist between ray-shooting for microlensing (Section
\ref{sec:rayshooting}) and the volume rendering technique we describe here,
they are fundamentally different algorithms.

\noindent \textbf{Outline:} The algorithm may be divided into the following
steps:

\begin{enumerate}
\item Obtain the input data cube.
\item Create a 2D grid of output pixels to be displayed.
\item Generate a corresponding grid of \textit{rays}, where each is defined by
  a position (initially the centre of the corresponding pixel), a direction
  (defined by the viewing transformation) and a colour (initially black).
\item Project each ray a small distance (the \textit{step size}) along its
  direction.
\item Determine which voxel each ray now resides in.
\item Retrieve the colour of the voxel from the data volume.
\item Use a specified \textit{transfer function} to combine the voxel colour
  with the current ray colour.
\item Repeat from step (iv) until all rays exit the data volume.
\item Output the final ray colours as the rendered image.
\end{enumerate}

\noindent \textbf{Analysis:} We interpret the steps in the above outline as
follows:
\begin{itemize}
\item Steps (ii) to (v) and (vii) are all transform operations.
\item Step (vi) is a gather operation.
\end{itemize}

All steps perform work scaling with the number of output pixels, $N_{\rm
pxls}$, indicating there are no algorithmic bottle-necks and thus
acceleration should be applied to the whole algorithm equally.

Given that the number of output pixels is likely to be large and scalable, we
should expect the transforms and the gather, with their $\mathcal{O}(1)$ depth
complexities, to parallelise perfectly on many-core hardware.

The outer loop of the algorithm, which marches rays through the volume until
they leave its bounds, involves some branching as different rays traverse
thicker or thinner parts of the arbitrarily-oriented cube. This will have a
negative impact on the performance of the algorithm on a SIMD architecture
like a GPU. However, if rays are ordered in such a way as to maintain 2D
locality between their positions, neighbouring threads will traverse similar
depths through the data cube, resulting in little divergence in their branch
paths and thus good performance on SIMD architectures.

The arithmetic intensity of each of the steps will typically be low (common
transfer functions can be as simple as taking the average or maximum), while
the complete algorithm requires $\mathcal{O}(N_{\rm pxls}N_d)$ memory reads,
$\mathcal{O}(N_{\rm pxls})$ memory writes and $\mathcal{O}(N_{\rm pxls}N_d)$ function
evaluations for an input data volume of side length $N_d$. This global
arithmetic intensity of $N_d:1:N_d\alpha$ indicates the algorithm is likely to
remain bandwidth-bound.

The use of bandwidth will depend primarily on the memory access patterns in
the gather step (the transform operations perform ideal contiguous memory
accesses). During each iteration of the algorithm, the rays will access an 
arbitrarily oriented plane of voxels within the data volume. Such a pattern
exhibits 3D spatial locality, presenting an opportunity to cache the memory
reads effectively and thus obtain near-peak bandwidth.

We conclude that the direct ray-casting volume rendering algorithm is a good
candidate for efficient implementation on many-core hardware, although, in the
absence of transfer functions with significant arithmetic intensity, the
algorithm is likely to remain limited by the available memory bandwidth.

\subsection{Pulsar time-series dedispersion}
\label{sec:dedispersion}
\textbf{Introduction:} Radio-telescopes observing pulsars produce time-series
data containing the 
pulse signal. Due to its passage through the interstellar medium, the pulse
signature gets delayed as a  function of frequency, resulting in a
`dispersing' of the data. The signal can be `dedispersed' by assuming a
frequency-dependent delay before summing the signals at each frequency. The
data are dedispersed using a number of trial dispersion measures (DMs), from
which the true DM of the signal is measured.

There are two principle dedispersion algorithms used in the literature: the
direct algorithm and the tree algorithm \citep{Taylor1974}. Here we consider the
direct method, which simply involves delaying and summing time series for a
range of  DMs. The calculation for each DM is entirely independent, presenting
an immediate opportunity for parallelisation. Further, each sample in the time
series is operated-on individually, hinting at additional fine-grained
parallelism.

\noindent \textbf{Outline:} Here we describe the key steps of the algorithm:

\begin{enumerate}
\item Obtain a set of input time series, one per frequency channel.
\item If necessary, transpose the input data to place it into channel-major order.
\item Impose a time delay on each channel by offsetting its starting location
  by the number of samples corresponding to the delay. The delay introduced
  into each channel is a quadratic function of its frequency and a linear
  function of the dispersion measure.
\item Sum aligned samples across every channel to produce a single
  accumulated time series.
\item Output the result and repeat (potentially in parallel) from step (iii) for
  each desired trial DM.
\end{enumerate}

\noindent \textbf{Analysis:} We interpret the above outline of the direct
dedispersion algorithm as follows:

\begin{itemize}
\item Step (ii) involves transposing the data, which is a form of
  \textit{gather}.
\item Step (iii) may be considered a set of \textit{gather} operations that
  shift the reading location of samples in each channel by an offset.
\item Step (iv) involves the summation of many time series. This is a nested
  operation, and may be
  interpreted as either a \textit{transform}, where the operation is to sum the
  time sample in each channel, or a \textit{reduce}, where the operation is
  to sum whole time series.
\end{itemize}

The algorithm therefore involves gather operations in addition to nested
transforms and reductions. For data consisting of $N_s$ samples for
each of $N_c$ channels, each step of the computation operates on all $\mathcal{O}(N_s
N_c)$ total samples. Acceleration should thus be applied equally to all parts
of the algorithm.

According to the depth complexity listed in Table
\ref{tbl:analysis_of_simple_algorithms}, the gather operation will parallelise
perfectly. The nested transform and reduce calculation may be parallelised in
three possible ways: a) by parallelising the transform, where $N_s$ parallel
threads each compute the sum of a single time sample over every channel
sequentially; b) by parallelising the reduce, where $N_c$ parallel threads
cooperate to sum each time sample in turn; or c) by parallelising both the
transform and the reduce, where $N_s \times N_c$ parallel threads cooperate to
complete the entire computation in parallel.

Analysing these three options, we see that they have depth complexities of
$\mathcal{O}(N_c)$, $\mathcal{O}(N_s\log{N_c})$ and $\mathcal{O}(\log{N_c})$
respectively. Option (c) would 
appear to provide the greatest speedup; however, it relies on using
significantly more parallel processors than the other options. It will in fact
only be the better choice in the case where the number of available parallel
processors is much greater than $N_s$. For hardware with fewer than $N_s$
parallel processors, option (a) will likely prove the better choice, as it is
expected to scale perfectly up to $N_s$ parallel threads, as opposed to the
less efficient scaling of option (c). In practice, the number of time samples
$N_s$ will generally far exceed the number of parallel processors, and thus
the algorithm can be expected to exhibit excellent parallel scaling using
option (a).

Turning now to the arithmetic intensity, we observe that the computation of a
single trial DM involves only an addition for each of the $N_s \times N_c$
total samples. This suggests the algorithm will be limited by memory
bandwidth. However, this does not take into account the fact that we wish to
compute many trial dispersion measures. The computation of $N_{\rm DM}$ trial
DMs still requires only $\mathcal{O}(N_s \times N_c)$ memory reads and
writes, but 
performs $N_{\rm DM} \times N_s \times N_c$ addition operations. The theoretical
global arithmetic intensity is therefore $1:1:N_{\rm DM}$. Given a typical
number of trial DMs of $\mathcal{O}(100)$, we conclude that the algorithm could, in
theory at least, make efficient use of all available arithmetic processing
power.

The ability to achieve such a high arithmetic intensity will depend on the
ability to keep data in fast memory for the duration of many arithmetic
calculations (i.e., the ability to efficiently cache the data). This in turn
will depend on the memory access patterns. We note that in general, similar
trial DMs will need to access similar areas of memory; i.e., the problem
exhibits some locality of reference. The exact memory access pattern is
non-trivial though, and a discussion of these details is outside the scope of
this work.

We conclude that the pulsar dedispersion algorithm would likely perform
to a high efficiency on a many-core architecture. While it is apparent that
some locality of reference exists within the algorithm's memory accesses,
optimal arithmetic intensity is unlikely to be observed without a 
thorough and problem-specific analysis of the memory access patterns.

\section{Discussion}
\label{sec:discussion}

The direct inverse ray-shooting method has been implemented on a GPU by
\citet{ThompsonEtal2010}. They simulated systems with up to $10^9$
lenses. Using a single GPU, they parallelised the interaction step of the
problem and obtained a speedup of $\mathcal{O}(100\times)$ relative to a
single CPU core 
-- a result consistent with the relative peak floating-point performance of
the two processing units\footnote{We note that \citet{ThompsonEtal2010} did
  not use the CPU's Streaming SIMD Extensions, which have the potential to
  provide a speed increase of up to $4\times$. However, our conclusion
  regarding the efficiency of the algorithm on the GPU remains unchanged by
  this fact.}.
These results validate our conclusion that the 
inverse ray-shooting algorithm is very well suited to many-core architectures
like GPUs.

Our conclusions regarding the pulsar dedispersion algorithm are validated
by a preliminary GPU implementation we have written. With only a simplistic
approach to memory caching, we have recorded a speedup of ~$15\times$ over an
efficient multi-core CPU code. This result is in line with the relative peak
memory bandwidth of the two architectures, supporting the conclusions of
Section \ref{sec:dedispersion} that, without a detailed investigation into the
memory access patterns, the problem will remain bandwidth-bound.

Some astronomy problems are well-suited to a many-core architecture, others
are not. It is important to know how to distinguish between these.
In the astronomy community, the majority of work with many-core hardware to
date has focused on the implementation or porting of specific 
codes perhaps best classified as `low-hanging fruit'. Not surprisingly,
these codes have achieved significant speed-ups, in line with the raw
performance benefits offered by their target hardware.

A more generalised use of `novel' computing architectures was undertaken 
by \citet{BrunnerEtal2007}, who, as a case study, implemented the two-point
angular correlation function for cosmological galaxy clustering 
on two different FPGA architectures\footnote{Field Programmable Gate Arrays
are another hardware architecture exhibiting significant fine-grained
parallelism, but their specific details lie outside the scope of this
paper.}.
While they successfully communicated the advantages offered by these new
technologies, their focus on implementation details for their FPGA
hardware inhibits the ability to generalise their findings to other
architectures.

It is interesting to note that previous work has in fact identified a number
of common concerns with respect to GPU implementations of astronomy 
algorithms. For example, the issues of
optimal use of the memory hierarchy and underuse of available hardware for
small particle counts have been discussed in the context of the direct N-body
problem (e.g., \citealt{BellemanEtal2008}). These concerns essentially
correspond to a combination of what we have referred to as memory access
patterns, arithmetic intensity and massive parallelism. While originally being
discussed as implementation issues specific to particular choices of software
and hardware, our abstractions re-cast them at the algorithm level, and
allow us to consider their impact across a variety of problems and
hardware architectures.

Using algorithm analysis techniques, we now have a basis for understanding
which astronomy algorithms will benefit most from many-core processors. Those
with well-defined memory access patterns and high arithmetic intensity stand
to receive the greatest performance boost, while problems that involve a
significant amount of decision-making may struggle to take advantage of the
available processing power.

For some astronomy problems, it may be important to look beyond the techniques
currently in use, as these will have been developed (and optimised) with
traditional CPU architectures in mind. Avenues of research could include, for
instance, using higher-order numerical schemes \citep{NitadoriMakino2008} or
choosing simplicity over efficiency by using brute-force methods (Bate
 et al. submitted). Some algorithms, such as histogram
generation, do not have a single obvious parallel implementation, and may
require problem-specific input during the analysis process.

In this work, we have discussed the future of astronomy computation,
highlighting the change to many-core processing that is likely to occur in
CPUs.

The shift in commodity hardware from serial to parallel processing units will
fundamentally change the landscape of computing. While the market is already
populated with multi-core chips, 
it is likely that chip designs will undergo further significant changes in
the coming years. We believe that for astronomy, a generalised methodology
based on the analysis of algorithms is a prudent approach to confronting these
changes -- one that will continue to be applicable across the range of
hardware architectures likely to appear in the coming years: CPUs, GPUs and
beyond.

\section*{Acknowledgments}
We would like to thank Amr Hassan and Matthew Bailes for useful discussions
regarding this paper, and the reviewer Gilles Civario for helpful
suggestions.

\bibliography{abbrevs,benbarsdell}{}

\begin{thebibliography}{}

\bibitem[\protect\citeauthoryear{{Amdahl}}{{Amdahl}}{1967}]{Amdahl1967}
{Amdahl} G.~M.,  1967, in AFIPS '67: Proceedings of the American Federation of
  Information Processing Societies Conference {Validity of the single processor
  approach to achieving large scale computing capabilities}.
pp 483--485

\bibitem[\protect\citeauthoryear{Asanovic, Bodik, Catanzaro, Gebis, Husbands,
  Keutzer, Patterson, Plishker, Shalf, Williams \& Yelick}{Asanovic
  et~al.}{2006}]{AsanovicEtal2006}
Asanovic K.,  Bodik R.,  Catanzaro B.~C.,  Gebis J.~J.,  Husbands P.,  Keutzer
  K.,  Patterson D.~A.,  Plishker W.~L.,  Shalf J.,  Williams S.~W.,    Yelick
  K.~A.,  2006, Technical Report UCB/EECS-2006-183, The Landscape of Parallel
  Computing Research: A View from Berkeley,
  \verb+http://www.eecs.berkeley.edu/Pubs/TechRpts/2006/EECS-2006-183.html+.
EECS Department, University of California, Berkeley

\bibitem[\protect\citeauthoryear{Asanovic, Bodik, Demmel, Keaveny, Keutzer,
  Kubiatowicz, Morgan, Patterson, Sen, Wawrzynek, Wessel \& Yelick}{Asanovic
  et~al.}{2009}]{AsanovicEtal2009}
Asanovic K.,  Bodik R.,  Demmel J.,  Keaveny T.,  Keutzer K.,  Kubiatowicz J.,
  Morgan N.,  Patterson D.,  Sen K.,  Wawrzynek J.,  Wessel D.,    Yelick K.,
  2009, Communications of the ACM, 52, 56

\bibitem[\protect\citeauthoryear{{Belleman}, {B{\'e}dorf} \& {Portegies
  Zwart}}{{Belleman} et~al.}{2008}]{BellemanEtal2008}
{Belleman} R.~G.,  {B{\'e}dorf} J.,    {Portegies Zwart} S.~F.,  2008, New
  Astronomy, 13, 103

\bibitem[\protect\citeauthoryear{Blelloch}{Blelloch}{1996}]{Blelloch1996}
Blelloch G.~E.,  1996, Commun. ACM, 39, 85

\bibitem[\protect\citeauthoryear{{Briggs}}{{Briggs}}{1995}]{Briggs1995}
{Briggs} D.,  1995, PhD thesis, New Mexico Institute of Mining and Technology

\bibitem[\protect\citeauthoryear{{Brunner}, {Kindratenko} \& {Myers}}{{Brunner}
  et~al.}{2007}]{BrunnerEtal2007}
{Brunner} R.~J.,  {Kindratenko} V.~V.,    {Myers} A.~D.,  2007, in NSTC '07:
  Proceedings of the NASA Science Technology Conference {Developing and
  Deploying Advanced Algorithms to Novel Supercomputing Hardware}

\bibitem[\protect\citeauthoryear{Che, Boyer, Meng, Tarjan, Sheaffer \&
  Skadron}{Che et~al.}{2008}]{CheEtal2008}
Che S.,  Boyer M.,  Meng J.,  Tarjan D.,  Sheaffer J.,    Skadron K.,  2008,
  Journal of Parallel and Distributed Computing, 68, 1370

\bibitem[\protect\citeauthoryear{{Clark}}{{Clark}}{1980}]{Clark1980}
{Clark} B.~G.,  1980, A\&A, 89, 377

\bibitem[\protect\citeauthoryear{{Hamada}, {Nitadori}, {Benkrid}, {Ohno},
  {Morimoto}, {Masada}, {Shibata}, {Oguri} \& {Taiji}}{{Hamada}
  et~al.}{2009}]{HamadaEtal2009}
{Hamada} T.,  {Nitadori} K.,  {Benkrid} K.,  {Ohno} Y.,  {Morimoto} G.,
  {Masada} T.,  {Shibata} Y.,  {Oguri} K.,    {Taiji} M.,  2009, Computer
  Science - Research and Development, 24, 21

\bibitem[\protect\citeauthoryear{{Harris}}{{Harris}}{2005}]{Harris2005}
{Harris} M.,  2005, GPU Gems 2 - Mapping Computational Concepts to GPUs.
Addison-Wesley Professional, pp 493--508

\bibitem[\protect\citeauthoryear{{Harris}}{{Harris}}{2007}]{Harris2007}
{Harris} M.,  2007, NVIDIA Developer Technology whitepaper, pp 1--38

\bibitem[\protect\citeauthoryear{{H{\"o}gbom}}{{H{\"o}gbom}}{1974}]{Hogbom1974}
{H{\"o}gbom} J.~A.,  1974, A\&AS, 15, 417

\bibitem[\protect\citeauthoryear{{Jonsson} \& {Primack}}{{Jonsson} \&
  {Primack}}{2009}]{JonssonPrimack2009}
{Jonsson} P.,  {Primack} J.,  2009, ArXiv e-prints

\bibitem[\protect\citeauthoryear{{Kayser}, {Refsdal} \& {Stabell}}{{Kayser}
  et~al.}{1986}]{KayserEtal1986}
{Kayser} R.,  {Refsdal} S.,    {Stabell} R.,  1986, A\&A, 166, 36

\bibitem[\protect\citeauthoryear{Knuth}{Knuth}{1998}]{Knuth1998}
Knuth D.~E.,  1998, The art of computer programming, 2nd edn.
Vol.~3, Addison-Wesley Longman Publishing Co., Boston, MA, USA

\bibitem[\protect\citeauthoryear{Levoy}{Levoy}{1990}]{Levoy1990}
Levoy M.,  1990, ACM Trans. Graph., 9, 245

\bibitem[\protect\citeauthoryear{{Moore}}{{Moore}}{1965}]{moore1965}
{Moore} G.~E.,  1965, Electronics, 38

\bibitem[\protect\citeauthoryear{{Nitadori} \& {Makino}}{{Nitadori} \&
  {Makino}}{2008}]{NitadoriMakino2008}
{Nitadori} K.,  {Makino} J.,  2008, New Astronomy, 13, 498

\bibitem[\protect\citeauthoryear{{Schive}, {Tsai} \& {Chiueh}}{{Schive}
  et~al.}{2010}]{SchiveEtal2010}
{Schive} H.,  {Tsai} Y.,    {Chiueh} T.,  2010, ApJ.~Supp., 186, 457

\bibitem[\protect\citeauthoryear{{Schneider} \& {Weiss}}{{Schneider} \&
  {Weiss}}{1986}]{SchneiderWeiss1986}
{Schneider} P.,  {Weiss} A.,  1986, A\&A, 164, 237

\bibitem[\protect\citeauthoryear{{Schneider} \& {Weiss}}{{Schneider} \&
  {Weiss}}{1987}]{SchneiderWeiss1987}
{Schneider} P.,  {Weiss} A.,  1987, A\&A, 171, 49

\bibitem[\protect\citeauthoryear{{Taylor}}{{Taylor}}{1974}]{Taylor1974}
{Taylor} J.~H.,  1974, A\&AS, 15, 367

\bibitem[\protect\citeauthoryear{{Thompson}, {Fluke}, {Barnes} \&
  {Barsdell}}{{Thompson} et~al.}{2010}]{ThompsonEtal2010}
{Thompson} A.~C.,  {Fluke} C.~J.,  {Barnes} D.~G.,    {Barsdell} B.~R.,  2010,
  New Astronomy, 15, 16

\bibitem[\protect\citeauthoryear{{Wambsganss}}{{Wambsganss}}{1990}]{Wambsganss%
1990}
{Wambsganss} J.,  1990, PhD thesis, Thesis Ludwig-Maximilians-Univ., Munich
  (Germany, F.~R.).~Fakult{\"a}t f{\"u}r Physik., (1990)

\bibitem[\protect\citeauthoryear{{Wambsganss}}{{Wambsganss}}{1999}]{Wambsganss%
1999}
{Wambsganss} J.,  1999, Journal of Computational and Applied Mathematics, 109,
  353

\bibitem[\protect\citeauthoryear{{Wayth}, {Greenhill} \& {Briggs}}{{Wayth}
  et~al.}{2009}]{WaythEtal2009}
{Wayth} R.~B.,  {Greenhill} L.~J.,    {Briggs} F.~H.,  2009,
  Pub.~Astron.~Soc.~Pacific, 121, 857

\end{thebibliography}
\bibliographystyle{mn2e}

\label{lastpage}

\end{document}